\input amstex
\documentstyle{amsppt}
\magnification\magstep1

\def\doublespace{\advance\baselineskip by 12pt}

\pagewidth{5in}
\pageheight{8in}

\hcorrection{.7pc}

\hfuzz1pc 
\loadbold

\hyphenation{amsppt}
\TagsOnRight
\CenteredTagsOnSplits

\define\AMS{American Mathematical Society}

\overfullrule=0pt

\define\thismonth{\ifcase\month 
  \or January\or February\or March\or April\or May\or June%
  \or July\or August\or September\or October\or November%
  \or December\fi}


\font\fourtn=cmr10 scaled \magstep2
\font\fourtnsy=cmsy10 scaled \magstep3
\font\fourtnbf=cmbx10 scaled \magstep3
\textfont2=\fourtnsy

\shipout\vbox to\vsize{%
\parindent=0pt 
\vskip5pc
\centerline{\fourtnbf  Finite Renormalization II*} 
\medskip
\centerline{\fourtnbf }
\bigskip
\medskip
\rightline{\fourtn  March 1999}
\vfill
\tenpoint

* Based on speculation concerning CODATA's delayed  1999-2000? Report.

\medskip
\begingroup\obeylines
\ \ \ \ \ \ \ \ \ \TeX{} is a trademark of the \AMS{}.
\endgroup
}


\topmatter
\title\nofrills  Finite Renormalization II*  \endtitle
\author { James G. Gilson} \endauthor
\affil { School of Mathematical Sciences,
             Queen Mary College University of London,
             Mile End Road,
             London E1 4NS,
             United Kingdom.}\endaffil
\address { School of Mathematical Sciences,
             Queen Mary College University of London,
             Mile End Road,
             London E1 4NS,
             United Kingdom.}\endaddress
\email {j.g.gilson\@qmul.ac.uk} \endemail


\thanks * See section 5 of this article \endthanks

\abstract
This paper is an updated version  of the paper of similar title published in September 1998$^{21}$ modified to take into account recent experimental results and recommendations from CODATA*$^{19}$ and also to incorporate a correction. The original abstract follows and is still valid.
A 1960's suggestion by R. P. Feynman, concerning the possibility of carrying out a {\it finite\/}
renormalization procedure in quantum electrodynamics, is here implemented using a newly discovered
{\it formula\/} for $\alpha$, the fine structure constant. 
 \endabstract

\endtopmatter
\document

\head 1 Introduction \endhead
Quantum electrodynamics, QED$^{1,2,3}$ is widely accepted as being the most successful theory of fundamental
processes that has been assembled up to the present time. Assembled is perhaps a better term
than discovered or created because it reflects it evolutionary development during much of this
century. Its success lies in its power to predict measurable characteristics of a wide and very
important range of physical systems. It is limited
essentially to the interactions between electronic systems and the electromagnetic field. It is
certainly difficult to overstate it significance in the general context of physical theory and
in the numerical accuracy of its predictions it is preeminent, giving some 
results to twelve places of decimals. It has been the main source of ideas as to how the more
general theories$^4$ necessary in the high energy context might be constructed. However, there is
a negative aspect of this success which stems from the very techniques that have been employed
in the assembling of QED. It is no exaggeration to remark that the development of the QED structure
has been plagued with infinities. To see how this has come about we shall now {\it very\/}
briefly describe the form of calculational algorithm that QED consists of at its present
stage of development.
The essential basic computational machine is a series expansion, the S-matrix$^3$, in terms of a small
numerical
parameter called the fine structure constant$^{5}$, $\alpha\approx 1/137$. The size of this parameter
is the essential ingredient that gives the series some {\it sort\/} of convergence. The coefficients
of this series are complicated integrals over  momentum space. These coefficients together with
their appropriate power of $\alpha$ are the mathematical representations of the various Feynman
diagrams giving information about some physical process involving a specific power of
$\alpha$. However, most of these coefficient  are complex additions of a variety of integrals over
momentum space and among the integrals are found quite a few that diverge to infinity in varying
degrees. Much of the past work in QED research has been into finding techniques for evaluating the
difficult
integrals and into finding ways of making sense of the fact that they occur within the general
QED structure. This has  been accomplished by the introduction of methods, largely due to the work
of F. J. Dyson$^{6}$, R.P. Feynman$^{2,4}$, and J. Schwinger$^{7}$ 
for the evaluation of the rogue integrals while leaving undisturbed any real physical significance
that they might have. This step is followed by {\it renormalization\/}, or the absorption
of the offending
parts into the fine structure constant power that goes along with that particular integral.
The effect of the renormalization procedure is to produce a renormalized series in powers
of the renormalized fine structure constant and with coefficients with no infinities and which
represent only recognizable physical effects.

The preceding description of the state of affairs is
exceeding over simplified but motivates a rather uncomplicated application of recent work$^8$
by the present author to the evolution of QED. Greatly detailed discussion of the complications
of renormalization can be found in references$^{1,3}$.  Thus it is that the great success of
QED which there is no wish to minimize is accompanied with the virulent question of the validity
of the mathematics process that neatly throws away infinities and incorporates them into the fine
structure constant which when used to generate the initial expansion was thought to be a definite  
numerical quantity approximately equal to $1/137$. This dilemma is almost certainly what Feynman
had in mind when he made a suggestion in a book$^{4}$ published in 1961. Essentially,
he suggested that if
a formula could be found for the fine structure constant in some future theory, not necessarily QED,
it would be likely that a renormalization program
could be carried through by {\it finite\/} steps. He was writing there specifically about charge
renormalization though clearly if such a procedure can be accomplished with regard to charge
renormalization its extension to the more general situation would likely follow without great difficulty.
Thus in this short article we shall concentrate on the question of finding a finite
renormalization scheme for charge. The result from QED that we require is the logarithmically
divergent
expression for
the $Z_3$ renormalization factor which is the factor multiplying the {\it theoretical\/} charge
{\it e\/} in
$$ e_R=\left (1 -{\alpha\over {3 \pi}}\ln{\Lambda ^2\over m^2}\right )^{1/2}e\tag 1.1$$
and is 
$$Z_3=\left (1 -{\alpha\over {3 \pi}}\ln{\Lambda ^2\over m^2}\right )^{1/2}.\tag 1.2$$
$e_R$ is the renormalized charge which will be taken to be the actual measured$^{9}$
physical value.
This simple formula is obtained from  QED by quite an elaborate series of steps and
involves the addition of  an infinite sum of Feynman diagrams as explained in references$^{1,2,3,4}$
The quantity
$\Lambda$ is the {\it cutoff\/} value for mass in the momentum space integrals and
would go to infinity
if the integral it is related to were fully evaluated according to the tenets of QED.
However, it is kept at some unspecified but finite value
greater than $m$ on the understanding that it should go to infinity when its integral is suitably
absorbed into the renormalized charge. Clearly it needs to be finite if sense is not to be violated
while manipulations take place using equation (1.1).
It should be emphasized that the result (1.1) arises from summing all relevant so
 called bubble graphs
so that in a sense it is an exact result and not just an approximation associated
with some power of $\alpha$ or some term in the S-matrix series expansion.
It is more convenient to use the fine
structure constant version of equation (1.1) which is obtained by squaring both sides of (1.1)
and then multiplying both sides 
by $1/(4 \pi \epsilon _0 \hbar c)$ by which means we obtain
$$ \alpha _R=\left (1 -{\alpha\over {3 \pi}}\ln{\Lambda ^2\over m^2}\right )\alpha\tag 1.3$$
with $\alpha _R$ being the renormalized fine structure constant and plain $\alpha$ being the
theoretical one. The value of the renormalized fine structure constant $\alpha _R$ will also
simultaneously
coincide with the value from the
physical definition,
$$\alpha _R = e_R^2/(4\pi \epsilon _0 \hbar c).\tag 1.4$$
This section will now be completed with my interpretation of
a perspective on the renormalization factors such as $Z_3$  for which I am
indebted to Professor C. W. Kilmister.  

{\it The introduction of the cutoff in the $Z_3$ multiplier is clever because it enables this
potentially or {\it  actually\/} divergent quantity to be {\it manipulated\/}.
However, manipulated with what could be a false sense of security. This is because $Z_3(\Lambda)$
is, as here indicated, defined as a function of $\Lambda$ which when convenient
is assumed to be finite  but which has no defined {\it definite\/} finite values within
the pattern of QED ideas. The definition of $Z_3(\Lambda )$ as a function of $\Lambda$
has the very restricted domain meaningful in QED of just $\Lambda$
larger than any finite quantity.
Further, there seems no prospect of assigning finite values to
the parameter $\Lambda$ for any QED originated reason. Thus some structure external to 
QED is needed in order to give some credence to an extension of the domain in which 
$\Lambda$ can vary and have definite numerical values. Only then will it be  possible
to identify, at the very least, a conceptual range in which the manipulations of
$Z_3(\Lambda)$ are meaningful. A related reason why  great care is needed in working with
such potentially divergent quantities is that any finite quantities that might
occur additively with them  will have the ambiguous status of being present
in some sense but numerically quenched and of uncertain relevance.}

In the following sections, it will be shown that full
account can be taken of this caveat.  
Equations (1.3) and (1.2) are  all that is needed from QED to give firstly
a very simple description of a finite renormalization scheme. Possible complications
that might be contemplated will be discussed in section 4.  With regard to the question of what Feynman {\it exactly\/} or {\it precisely\/} meant by his rather enigmatic remarks in the sixties suggesting the possibility of finite renormalization consequent on the finding of a {\it formula\/} for $\alpha$ we  cannot today be entirely certain.  However, in section 3 of this paper a finite approach  possibility for the electrical {\it charge\/} problem  is demonstrated in detail and  seemingly  conforms closely to what he apparently thought would follow the discovery of a formula for $\alpha$. That it has now been possible to construct a {\it finite\/} renormalization  scheme seems to strongly confirm the correctness of Feynman's conjecture and also the soundness of  his intuition.    

 \head 2 The formula for $\alpha$\endhead
The present author$^9$ has recently found, from theory$^{8,10,11}$ a QED independent formula for the
fine structure constant
which depends on two integer parameters which are denoted by $N$ and $N_b$. The formula is
$$\alpha (N,N_b) = N_b \cos (\pi /N)\tan (\pi/(NN_b))/\pi.\tag 2.1$$
This formula arises from the authors alternative theory$^{10,11}$ for the quantum process
and is entirely independent from the QED structure This alternative theory will be referred
to as GT in the following work. GT has a Schr\"odinger equation and special relativity$^{12}$
basis. The {\it integer\/} $137$ plays a special role in this theory so that for this and
other reasons  GT has some
common conceptual ground with the work of Eddington$^{13}$ and the more recent works of
Bastin$^{15,16}$ and Kilmister$^{14,17}$ on the Combinatorial Hierarchy$^{16}$.
The general validity of the formula (2.1) has now been strongly reinforced by showing that it has a fundamental and inevitable significance for the structure of the first Bohr orbits of the whole family of hydrogen like atoms$^{20}$.
We shall here employ the one variable subfunction,
$$\alpha (N_b)=\alpha (137,N_b)= N_b \cos (\pi /137)\tan (\pi/(137N_b))/\pi ,\tag 2.2$$
in which $N$ is kept fixed at the value $137$ leaving only $N_b$ to range over
positive integral values. $\alpha (N_b)$  or some multiple of it will,
 for the purposes of this article, be taken to be our
{\it theoretical\/} fine structure constant. Other than the fact that it is now a variable
quantity depending on $N_b$ it corresponds with the plain $\alpha $ used earlier and it is
to be regarded as the initial
small parameter expander or unrenormalized $\alpha$ that is used in the $S$-matrix series.
Also in that context the parameter $N_b$. would be expected to have some definite
integral value from its allowable range. The choice of $N_b$ used in the initial expansion
can be
regarded as representing a decision as to how well we know the physical coupling
before the S-matrix complications are turned on or, in other words, a selection of
a {\it less dressed\/} coupling constant.

The 1999-2000? CODATA\footnote"*"{See section 5 of this article} recommended value  and reliability range for $alpha$ is substantially different from the 1986 value, the new range being {\it outside\/} the old.    In table 1 below the values of $\alpha (N_b)$ are given
that lie
in the experimental and recommended CODATA$^{9}$ 1986 range range $\alpha _{Nucmin}$ to $\alpha _{Nucmax}$ with
center at the best experimental value $\alpha _{Nuc}$. The nearest value from GT taken by
the function $\alpha (N_b)$ to
this center value is given by taking $N_b=25$ and this value differs from the center value
by approximately $2.3\  10^{-11}$. I have decided to leave this information intact but {\it italicised\/} in this version II* of this article for ease of reader reference and as an illustration of the nature of the predictive character of the formula and its dependence on having definite {\it experimental\/} information concerning the range of the possible values.  What was thought to be definite information concerning range is now likely to have  changed as a result of the forthcoming CODATA* report and speculation based on {\it rumour\/} concerning the contents of this yet unpublished report  follows after the paragraph involving the old $\alpha _b$, equation (2.3).

\newpage
{\it  
\noindent {\bf Table 1}\phantom{aaaaaaaaaaa} $	N_b$ solutions  with 1986 CODATA $\alpha$ values and bounds 
\settabs 2 \columns
\+ $\phantom{aaaaaaaaaaa}N_b$&\phantom{xxxxxxxxxxx}$\alpha$-values\cr
\+ $\phantom{aaaaaaaaaaa}\ $&$\alpha _{Nucmax}=0.007297353410$\cr
\+ $\phantom{aaaaaaaaaaa}24$&$\alpha (24)\ \ \ \ =0.007297353232$\cr
\+ $\phantom{aaaaaaaaaaa}\ $&$\alpha _{Nuc\phantom{max}}=0.007297353080$\cr
\+ $\phantom{aaaaaaaaaaa}25$&$\alpha (25)\ \ \ \ =0.007297353057$\cr
\+ $\phantom{aaaaaaaaaaa}26$&$\alpha (26)\ \ \ \ =0.007297352903$\cr
\+ $\phantom{aaaaaaaaaaa}27$&$\alpha (27)\ \ \ \ =0.007297352766$\cr
\+ $\phantom{aaaaaaaaaaa}\ $&$\alpha _{Nucmin}=0.007297352750$\cr

The next nearest value to the center value $\alpha _{Nuc}$ is given by taking $N_b=24$ but
this differs from the center value by approximately $15.2\ 10^{-11}$.
Thus taking the simplistic but possible reasonable view that the center of the range
is better than the more removed parts, $N_b=25$ is better than $N_b=24$
by a factor of about $6$. There is another reason$^{16}$ that $25$ might be preferred to $24$
and that is because $25$ has the simple relation $25=k_2^2/4$ with the second combinatorial$^{16}$
special number  $k_2=10$, the $4$ relating to the value of the total solid angle $4\pi$.
However, this connection may be dismissed as pure {\it numerology\/}\ ! 
Taking account of the latest measured and theoretically adjusted value$^{9}$ ascribed to
the fine structure constant, the best value from GT for the fully dressed coupling
constant is obtained from taking $N_b=25$
and is given by,
$$\alpha_b=\alpha (25)=0.007297353057\tag 2.3$$
to twelve places of decimals. It is emphasized that the value of $\alpha (N_b)$ at $N_b=25$ given in
(2.3) is to be regarded as the value obtained by measurement
and corresponds to the renormalized $\alpha _R$ used earlier.}

The next CODATA\footnote"*"{See section 5 of this article} report$^{19}$ is likely to gives a recommended value for $\alpha$
$$\alpha=0.007297352534(13)^*.$$ In contrast with the 1986  range,  there is only one value given by the predictive formula (2.2) that lies in this new 1999-2000? CODATA* range and that is the value when $N_b=29$ which is
$$\alpha _b=\alpha(29)=0.007297352532.\tag 2.3b$$
This prediction differs from the {\it speculated\/} recommended value by approximately  that 2 parts in $10^{12}$ parts. This  is certainly very impressive accuracy assuming the speculation is correct. Thus from now on in this version of the article we shall use the parameter value $N_b=29$.

From (2.2) and (2.3b) it follows that
$$\alpha _b ={29\tan (\pi /(137\  29))\over {N_b\tan (\pi /(137 N_b))}}\alpha (N_b).\tag 2.4$$
Thus from the theoretically deduced $\alpha (N_b)$ value we find  {\it theoretically\/}
deduced from GT a charge
renormalization factor $Z_3$ of form,
$$Z_3=\left ({29\tan (\pi /(137\  29))\over {N_b\tan (\pi /(137 N_b))}}\right )^{1/2}\tag 2.5$$
the square of which by multiplication converts the theoretical value of $\alpha$ to the measured
or renormalized value.
We note that the renormalization factor from QED given by (1.3) depends on the
cutoff parameter mass
$\Delta $ from  the momentum space integrations. There is no compelling reason for
this to be infinite
other than for the reason that the QED formalism imposes no restriction on the integrations over
momentum space. If there were such restriction in QED then the nature of the problem would
change drastically. However, all the evidence up to date is that the possibly infinite integrals
do not contribute to the physical information contained in that theory. This is the reason
that the possible infinities can be dumped into the renormalized coupling constant and in effect
disregarded. Of course, the conceptual difficulties are not removed by recognizing this aspect.
Thus it would be quite acceptable and rather convenient if the finite cutoff $\Delta$ used
in QED as an infinity remedial was in fact an {\it actually\/} finite limit for some hitherto
unnoticed reason.
As the possible infinite integrals are physically irrelevant anyway they could still be dumped
into a renormalized coupling constant but conceptually rather more easily.
This is the possibility that Feynman
had in mind but then he had no theoretical formula for the fine structure constant.
If we consider the theoretically deduced renormalization factor $Z_3$ from GT as given by (2.5),
we see that it also
contains a cutoff quantity, the value of $N_b$ at the specific value $N_b=29$.
The cutoff in GT arises from what seems to be an upper limit to the snap$^8$ bending of a wave
in order for it to make a best fit to a curved contour along which it is moving.
The formula (2.5) still makes
sense for values above $N_b=29$ but there seems to be this physical cutoff at the actual
measured value of the fine structure constant. Thus we are motivated to investigate the possibility
that the two cutoffs, the one from QED and the one from GT$^{10,11,18}$,
a theory external to QED, are related
functionally and in some physically determined sense.

\head 3 Finite Renormalization \endhead
It is possible to exploit the fact that the QED $Z_3(\Lambda)$ is a function of $\Lambda$ by
equating the two expressions (1.2) and (2.5) for the $Z_3$ charge renormalization
factors. Making this step we get the relation
$$\left (1 -{\alpha (N_b) \over {3 \pi}}\ln{\Lambda ^2 (N_b)\over m^2}\right )=
\left ({29\tan (\pi /(137\  29))\over {N_b\tan (\pi /(137 N_b))}}\right )\tag 3.1$$
where now the $\alpha$ in the QED factor (1.2) has been identified with the theoretical
fine structure constant and the possible dependence of the QED cutoff, $\Lambda$, on $N_b$
has also been taken into account.
We observe that when $N_b=29$, $\Lambda =m$ as then both sides of the equation reduce to unity
making the renormalized charge and theoretical charges equal in both QED and in GT.
Solving equation (3.1) for $\Lambda (N_b)$, we obtain,
$$\Lambda (N_b)=m\exp \left ({ 3\pi\over {2 \alpha (N_b)}}{\left (1- {29\tan
 (\pi /(137\  29)\over {N_b\tan (\pi /(137 N_b))}}\right )}\right ).\tag 3.2 $$
The $Z_3(N_b)$ from 	GT is a definite and meaningful function of $N_b$. Thus identifying
the two different $Z_3$ factors from QED and GT and the consequent generation
 of the definite and finite
relation (3.2) between $N_b$ and $\Lambda (N_b)$ adds to QED from the external system GT
the possibility for rationally ascribing values to $\Lambda (N_b)$ other than just infinity.
It is then possible to make sensible a QED $Z_3(\Lambda)$ functional dependence on
an extended domain of $\Lambda$ values. Thus effective account is taken of the Kilmister caveat.
The relation between the two cutoff parameters given in equation (3.2) solves the problem
of constructing a finite scheme for charge renormalization as originally conceived by
R.P. Feynman. The prescription for its use is thus as follows:-

{\it Expand the S-matrix as usual
using a value for the fine structure constant in its theoretical form $\alpha (N_b)$
at a specific value of $N_b$.
Cutoff the logarithmic divergent integrals arising in the coefficients at the corresponding
$N_b$ value as given by formula (3.2). Dump these unwanted finite integrals into the renormalized
$\alpha _R=\alpha _b$ as given by equation (1.3). Further, if the $N_b$ value chosen at the
outset is $29$
then all the potentially logarithically divergent integrals will evaluate to zero because
$\Lambda (N_b)=m$ in this case and so all such integrals can be ignored anyway. In all 
these situations the final S-matrix expansion will be in terms of the renormalized measured
fine structure constant and will have only finite integral coefficients.\/}
 \head 4 Complications \endhead
It has been shown how the charge renormalization process that plays the central role
in QED of separating physically relevant information from possible divergent but otherwise
redundant information can can be carried through by finite steps according to a suggestion
by R.P. Feynman.  As we have noted, $Z_3$ is usually regarded as being in some sense a
divergent quantity. Thus it might be argued that because its derivation in QED
involves the loss of small quantities that would be  important when in fact $\Lambda$ is small
as in the case in hand, the formula (1.1) for $Z_3$ may not be adequate in its simple form.
This issue is another aspect of the Kilmister caveat. It will now be shown
that a more general version of (1.1)
covering all such possible
small quantity omissions which has been obtained from QED can be used in place of (1.2).
This more general $Z_3$ has the form
$$ Z_3=\left (1 -{\alpha\over {3 \pi}}(D({\Lambda \over m})+G(\alpha ))\right )^{1/2}.\tag 4.1$$
QED derived versions for the functions $D(x)$ and $G(x)$ are
$$D(x)=\ln (x^2+1) +{1\over x^2 + 1} -1\tag 4.2$$
and
$$G(x)=-2/3.\tag 4.3$$
The value given here at (4.3) for the constant G(x) is the correction referred to in the abstract. The original value used was $5/6$. This change in magnitude and sign makes no difference to the question of the validity of the idea of how to handle such constant terms that might have been missed in the QED renormalization argument. However, it does make some difference to {\it numerical\/} values of derived quantities such as $\lambda$ for example in equation (4.10).  The wrong number was used earlier as a result of this author misinterpreting numerical results from published work on QED.
If we use these two function in (4.1) we find that the quantity
$$C=-{\alpha\over {3 \pi}}(D({\Lambda \over m})+G(\alpha ))\tag 4.4$$
does not have the value zero when ${\Lambda \over m}=1$ the value at which the potentially
divergent integral actually evaluates to the value zero corresponding with the  renormalized and
theoretical value of the fine structure constant coinciding. This rather satisfactory behavior
associated with the original $C^{\prime}=-{\alpha\over {3 \pi}}\ln{\Lambda ^2\over m^2}$ in (1.2)
 can be restored for the renormalization factor (4.1)
by making use of some freedom of choice that we have in selecting the theoretical fine structure
constant which in this more general situation we shall denote by $\alpha _g(N_b)$. Clearly
there is freedom in this respect because {\it mathematically\/} any small value could
in principle be used to generate the original S-matrix expansion. In keeping with
the usual thinking about the theoretical $\alpha _g$, although here it will be finite,
it is still not a necessarily a
{\it physically\/} measurable quantity.
However, it is very important that the renormalized $\alpha _R$ which is {\it physically\/}
measurable should
have the most accurate numerical value attached that has been agreed from experimentation$^{9}$.
Thus in the more general situation we define the theoretical fine structure constant
$\alpha _g(N_b)$ by
$$\alpha _g(N_b)=\lambda \alpha (N_b),\tag 4.5$$
where the multiplier $\lambda$ is chosen so that also in the more general situation
$$N_b=29\implies{\Lambda \over m}=1 \implies \alpha _R = \alpha (29)\tag 4.6$$ and $\alpha (N_b)$
is still given by the function defined in (2.2).
Thus the new $Z_3$ factor from GT will be given by,
$$Z_3^2={\alpha _R \over \alpha _g}={\alpha (29)\over {\lambda \alpha (N_b)}}=
    \left ({29\tan (\pi /(137\  29))}\over {\lambda N_b\tan (\pi /(137 N_b))}   \right ).\tag 4.7$$ 
The relation between the two cutoff parameter $\Lambda$ and $N_b$ is now obtained by 
equating (4.1) squared  with (4.7) and in this more general representation becomes
$$D({\Lambda (N_b)\over m})={3\pi\over{\lambda\alpha (N_b)} }{\left (1-\left ({29\tan (\pi /(137\  29))
\over {\lambda N_b\tan (\pi /(137 N_b))}}\right )\right )} - G(\alpha).\tag 4.8$$
Applying the conditions (4.6) in equation (4.8) gives a quadratic equation for $\lambda$
with solutions
$$\lambda={2\over{1\pm\left (1 - {4 \alpha _R\over {3 \pi} }(\ln (2) + {7\over 6})\right )^{1/2}}}.\tag 4.9$$
The value of $\lambda$ given by the positive sign in (4.9) is near to unity. The value with the
negative sign is $\lambda \approx -2728.52$. So that to keep to a value for the theoretical fine
structure constant, $\alpha _g (N_b)=
\lambda \alpha (N_b)$, near to
$1/137$ the positive sign solution is chosen. Thus giving 
 the value of $\lambda$ as
$$\lambda \approx {1\over {1-{\alpha _R \over 3\pi}(\ln (2) - {7\over 6})}}\approx 0.999633635.\tag 4.10$$
With the value (4.10) for $\lambda$ the cutoff quantity $\Delta (20)$, evaluated at $N_b=20$, assumes the value $\Delta (20)=1.000594024m$ just slightly greater than the electronic rest mass. 
The intrinsic formula (4.8) for $\Lambda (N_b)$ in terms of $N_b$ is not quite as simple as
(3.2) but it will still give the cutoff in $\Lambda (N_b)$ that arises from the values of
$N_b$ and apart from being a little more complicated the finite renormalization
scheme is still effective.
\head 5 Prediction of  $\alpha$'s measured value\endhead
In this section, additional to the original version of this paper$^{21}$, it is explained why the change from $N_b=25$ to $N_b=29$ has become necessary in the light of  long awaited next CODATA report on the measured values of the fundamental constants. This report is some years overdue and it is still uncertain whether it will appear this year  1999 or in the year 2000. In the light of this uncertainty, I have decided to anticipate its findings for the experimental  numerical value of $\alpha$ by speculation based on rumour.
 
Firstly, we consider my theoretical formula  for $\alpha$ and the nature and significance of the prediction involved in its discovery. The formula (2.1) in question is
$$\alpha (N,N_b)= N_b \cos (\pi /N)\tan (\pi/(N_b \times N))/\pi.\eqno 5.1$$
The two dimensional domain of this function is taken to be all positive integer pairs $(N,N_b)$.
The value ascribed to the integer $N$ is dominant in determining the function value and  is involved as the angle $\chi =\pi /N$ which is the size of the angular sector that a trapped electron  wave would occupy in a circular Bohr orbit if it were moving with the quantized velocity $N\alpha c$ with $N=137$. The second integer $N_b$ has a more subtle significance and its value controls very small variations from the value determined by $N$. It measures the number of {\it linear\/} quantized subdivisions in the trapped wave that occur  so that the wave  can aligns itself as near as possible  to a circular section of its orbit while its mean velocity  coincides with the quantized velocity associated with the orbit. In a sense this is the requirement that the relativistic contraction factor  for the whole wave body should  have the usual relativistic dependence on the quantized velocity in orbit. However,
this conformity imposes no restriction on the actual integral value of $N_b$. The requirement of this conformity is simply a result of the quantization of the wave length into a finite number of linear segments.  Thus the formula is predictive in that it says the value of the fine structure constant is determined by two integers but the two integers are {\it not\/} given by the theory and do have to be determined by detailed experimental knowledge of the range of values that the fine structure constant is considered to lie within. The integer value $N=137$ is however inevitable as any change in this would give values for $\alpha$ greatly outside the range of values generally accepted as inferred from  experimental measurements. The choice of the integer pair has to be made from the experimental information concerning the measured range of values that  is assessed as contained the numerical value of  $\alpha$ as near to certainty as possible. Thus if this range is large there are many possibilities and if the range is sufficiently small the possibilities could be reduced to a single unique value. The 1986 CODATA$^{9}$ range within which the numerical value of $\alpha$ was claimed to reside was small but the predictive function (5.1) had four possibilities $N_b=24,25,26,27$ with 25 the value n
earest to the centre as shown in table 1. Thus it seemed that the value $N_b=25$ would give the best prediction and this was reinforced by the fact that the physical definition for $\alpha$ equation (1.4)  gave a value only differing from this value by approximately $2$ part in $10^{11}$. Thus $N_b=25$ seemed an excellent option and  this option was  used by the author in $^{21}$ on the fine structure constant.  However, CODATA is expected to publish its next report in $1999-2000$ and there are what can only be described as rumours that substantial changes are expected with possibly the new range being {\it outside\/} the old range.

Firstly let us dwell on the way the CODATA published information and recommendations for fundamental constant values is constructed out of the enormous amount and very diverse world wide  experimental information they gather on the measured values of the various physical constants.  A {\it greatly\/} simplified description of this very complex operation is as  follows.  As far as the fine structure constant was concerned, in the 1996 report there were eight main largely independent different type of experimental information sources each source having its own values and reliability range. The central value given by the various sources were very disparate with values for the inverse fine structure constant ranging from $137.0359$ to $137.036$ roughly.  This is a large variation when we consider that we are attempting to get the error for $\alpha$ itself  down to less than something of the order of   $2$ parts in $10^{12}$ parts. The experimental information that has been gathered  for the next report is likely to be at least as widely varied as the earlier $1986$ version. Thus the daunting
problem for CODATA is firstly deciding on the comparative reliability and significance of the various sources.  Then a very elaborate  statistical analysis of the information to hand will hopefully generate a best value preferably the actual physical value for the fine structure constant together with a single range about this best value within which they can claim the true physical value is certain to lie. Whereas the final single {\it range\/} deduced by this analysis may be {\it very\/} reliable or even certain  it seems unlikely that their central value will be the true physical value unless they are very lucky. The new CODATA report  will have successfully had to perform  this task and have had to come up with values that any predictions can be checked against.  So that the coming report holds an exciting prospect for checking the accuracy of my predictive formula. As mentioned earlier, there are only rumours available at this moment of time. The hottest rumour at the moment 
of the date on this article  is that the new range is much smaller than the old
range and that the new recommended value for $\alpha$ is much nearer to $1/137.035999$ than was the old value.  If  this turns out to be the case, the value $N_b=29$ could be very near the mark and hopefully it might be the only value predicted within the new range. This would strongly confirm the validity of formula (5.1).  Such changes that might occur make no difference to questions of the validity or make changes of principle necessary in the theoretical  structure  that the formula was generated  from. However,  numerical values that have been obtained for other physical quantities in  reference$^{21}$ will be slightly changed as these  depend on the value of  $N_b$ that is supposed to give the actual physical ,{\it renormalized \/} value. This aspect will effect the author's application of the formula (5.1) to the problem of constructing a scheme of {\it finite\/} renormalization$^2$. Very small numerical changes in renormalized quantities will occur if  it is found necessary to replace $N_b=25$ with $N_b=29$. In particular, the physical or renormalized value of the fine structure constant will be given by $\alpha (137,29)$ rather than by  $\alpha (137,25)$. These changes have now been made in this version of the original paper.

\head 6 Conclusions\endhead
The conclusion regarding the {\it definitely\/}
finite terms that might be regarded as missing from the simple $Z_3$ formula (1.1) is that they can
easily be absorbed into the {\it bare\/} or theoretical fine structure constant.
That is into initial S-matrix expander for which there is {\it input\/} freedom of choice
except that it should have a value near to $1/137$. Thus because of this freedom
in the selection of the {\it finite\/}  theoretical fine structure constant  
our initial simple derivation
of the {\it finite\/} renormalization scheme in section 3 is not significantly unsound.  
Feynman's remarks in reference$^{4}$ were only about charge renormalization and here
also only charge renormalization has been considered so far.
In the more general divergency context, mass renormalization which as is well known can be
carried through to all orders in the coupling constant, should be shown to fit into a finite
renormalization scheme. In view of the ease with which {\it finite\/} charge renormalization
has been accomplished by using the new formula (2.1) for the fine structure constant
there seems no obvious
reason why a {\it finite\/} mass renormalization scheme should not also be constructed.
It is immediately obvious that if the cutoff in Feynman's mass counter term $\delta m=m_R-m$
is taken
to be the same quantity as the cutoff parameter $\Lambda$ used earlier and the formula for the
theoretical fine structure constant (2.2) is used in its representation, we have
$${\delta m (N_b)\over m}={\alpha (N_b)\over {2 \pi}}\left ({3\over 2}\ln 
({\Lambda ^2(N_b)\over m^2}) +
{3\over 4}\right ).\tag 6.1$$
This mass counter term is finite for positive integral values of $N_b$ and in particular
for the physical value $N_b=29$. Thus it can be manipulated without ambiguity in the
renormalization operations. We note the physical and finite value
$${\delta m (29) \over m}= {3 \alpha (29)\over {8\pi}}\approx 0.00089106.\tag 6.2$$
if (3.2) is used. As with the $Z_3$ factor, Feynman's simple version of
${\delta m (N_b)\over m}$ may
not be considered adequate when $\Lambda$ is finite. However, this is also easily generalized.
The main conclusion can be expressed as follows. An additional rule, equation (3.2) or
more generally (4.8), from
GT can be added to the QED
formalism restricting the integration range of the divergent integrals so that
the renormalization process can be made finite and therefore conceptually rational. The
renormalization process can then be seen as a {\it very\/} complex {\it one\/} step iteration
operation with only finite quantities.
A numerically exact (unrenormalized) fine structure constant is the initial input
expansion parameter with some reasonable finite starting value.
The whole QED formalism then, through the S-matrix expansion, generates a finite physically
exact fine structure constant to replace the initial parameter value by a single iteration
step under the control of an {\it externally\/} imposed constraint.
 \vskip 0.1cm
\noindent Acknowledgments: 
I am very grateful to Dr. P. J. McCarthy for  helpful
discussions on various aspects of the important part played by the fine structure constant in
fundamental physical theory. I am very grateful to  Professor C. W. Kilmister for similar
discussions and also for a number of significant suggestions along the unfolding of
this research project.    


\Refs
\refstyle{A}
\widestnumber\key{14}
\ref\key 1 \by Thirring, W.E. 1958 \book Principles of Quantum
Electrodynamics\publ Academic Press Inc.\endref
\ref\key 2 \by Feynman, R. P.1961\book Quantun Electrodyanamics, Reprint Series
\publ W.A. Benjamin,Inc,.New York.\endref
\ref\key 3 \by Bogoliubov N. N., Shirkoov, D. V. 1959\book Introduction to the Theory of
Quantized Fields Volume III \publ Interscience New York.\endref 
\ref\key 4 \by Feynman, R. P. 1961\book Theory of Fundamental Processes, Reprint Series
\publ W.A. Benjamin, Inc, New York.  \endref
\ref\key 5\by Sommerfeld, A. 1916\jour Annalen der Physik\vol 51{\rm , 1}\endref
\ref\key 6 \by F. J. Dyson 1949\paper The S-Matrix in Quantum Electrodynamics
\jour Phys. Rev.\vol 75\pages 1736\endref
\ref\key 7 \by Schwinger, J. 1953\paper The Theory of Quantized Fields II
\jour Phys. Rev.\vol 91\pages 713\endref
\ref\key 8 \by Gilson, J.G. 1996\paper Calculating the fine structure constant
\jour Physics Essays, Vol. 9 \issue 2 June\pages 342-353 \endref
\ref\key 9 \by Cohen, E. R. and Taylor, B. N. 1993\paper The Fundamental Physical Constants
\jour Physics Today (August)\pages BG9-15\endref
\ref\key 10\by Gilson, J.G. 1991\paper Oscillations of a Polarizable Vacuum
\jour Journal of Applied Mathematics and Stochastic Analysis
\vol 4\issue 11\page 95--110\endref
\ref\key 11 \by Gilson, J.G. 1994\paper Vacuum Polarisation and The Fine Structure Constant\jour
Speculations in Science and Technology \vol 17\issue 3 \pages 201-204\endref 
\ref\key 12 \by Rindler, W. 1969 \book Essential Relativity
\publ Van Nostrand Reinhold Company \endref 
\ref\key 13\by Eddington, A.S. 1946\book Fundamental Theory\publ Cambridge University Press\endref
\ref\key 14\by Kilmister, C.W. 1992\jour Philosophica\vol 50\pages 55 \endref
\ref\key 15\by Bastin, T. 1966\jour Philosophica Gandensia\vol 4\pages 77\endref 
\ref\key 16 \by Bastin, T., Kilmister, C. W. 1995 \book Combinatorial Physics
\publ World Scientific Publishing Co. Pte. Ltd.\endref
\ref\key 17 \by Kilmister, C. W. 1994 \book Eddington's search for a Fundamental Theory
\publ Cambridge University Press\endref
\ref\key 18 \by Gilson, J.G. 1997 \paper  Relativistic Wave Packing and  Quantization\jour
Speculations in Science and Technology \vol 20\issue 1  March\pages 21-31\endref
\ref\key19 \by  CODATA Report 1999-2000?\paper The Fundamental Physical Constants
\jour Journal of Physical and
Chemical Reference Data\endref
\ref\key20 \by  Gilson, J. G. 1998\paper Relativistic Length Contraction and Quantization
\jour Physical Interpretations of Relativity Theory, British Society for the Philosophy of Science Conference Proceedings,  Imperial College September\endref
\ref\key21 \by  Gilson, J. G. 1998\paper Finite Renormalization
\jour Speculations in Science and Technology, Vol 21, No 3,  September\endref


 
\endRefs
\enddocument